\def\be{\begin{equation}}
\def\te{\end{equation}}
\def\bea{\begin{eqnarray}}
\def\nn{\nonumber  \\}
\def\tea{\end{eqnarray}}

\documentclass[12pt]{article}
\baselinestretch
\begin{document}
\input epsf
\title{Qubit Decoherence and Non-Markovian Dynamics \\ at Low Temperatures via an
Effective Spin-Boson Model}
\author{K. Shiokawa
\thanks {E-mail address: kshiok@physics.umd.edu}
and B. L. Hu
\thanks{E-mail address: hub@physics.umd.edu}
\\ {\small Department of Physics, University of Maryland,
College Park, MD 20742, USA} }
\maketitle
\begin{abstract}
Quantum Brownian oscillator model (QBM), in the Fock-space
representation, can be viewed as a multi-level spin-boson model.
At sufficiently low temperature, the oscillator degrees of freedom
are dynamically reduced to the lowest two levels and the system
behaves effectively as a two-level (E2L) spin-boson model (SBM) in
this limit. We discuss the physical mechanism of level reduction
and analyze the behavior of E2L-SBM from the QBM solutions. The
availability of close solutions for the QBM enables us to study
the non-Markovian features of decoherence and leakage in a SBM in
the non-perturbative regime (e.g. without invoking the Born
approximation) in better details than before. Our result captures
very well the characteristic non-Markovian short time low
temperature behavior common in many models.
\end{abstract}
%
%
\newpage
\section{Introduction}

Recent development in quantum information processing and quantum
computation has attracted much attention to the study of discrete
quantum systems with finite degrees of freedom. The most commonly
used model is an array of interacting two-level systems (2LS) each
of which representing a qubit. As the system almost always
interacts with its environment, quantum decoherence in the system
usually is the most serious obstacle of actual implementation of
quantum information processors \cite{Dec96,PazZurek00,Unruh95}.
For this reason a detailed understanding of quantum decoherence in
open systems is crucial. There are a handful models  useful for
such studies, the  quantum Brownian motion(QBM)
\cite{Weiss99,QBM0,QBM1,ZHP93} is one, the spin-boson model
\cite{Weiss99,LCDFGZ87} is another: the system in the former case
is a harmonic oscillator and in the latter case a 2LS, both
interacting with an environment of a harmonic oscillator bath
(HOB).

Most qubit models presently employed are the results of picking
out the levels most relevant to the description of the qubit from
a multi-level structure. In atom optics, internal electronic
excitations are often approximated by a 2LS consisting of the
ground state and the excited state. A similar model is used for
the study of low temperature tunneling process where the two
levels degrees of freedom represent the quasi ground states in a
double well potential. The simplification to two-levels allows for
detailed analytical or numerical treatment, but this remains an
approximation applicable only when the effect of higher levels are
negligible, e.g. , at low enough temperature when higher levels
are not populated. However, in the presence of gate operation, the
existence of higher levels causes a leakage of the 2LS due to
transitions to other levels. Some extra perturbation may be
necessary to select or restrict multi-level structure into the
particular levels of interest\cite{TianLloyd00}. In order to make
a quantitative estimation of decoherence with a leakage effect, it
is more desirable to study open system models which maintain the
multi-level structure.

In the present paper, we study the aspects of realistic qubits
naturally arising from QBM, taking advantage of a fairly good
understanding from the detailed studies over the last few decades.
In particular, we focus on harmonic QBM, which can be viewed as an
$\infty$-level spin-boson model. Commonly used two-level
spin-boson model can be obtained by restricting the harmonic
oscillator Fock space to the lowest two levels. This
correspondence allows for a detailed analysis of the spin-boson
model from the known results of QBM. In particular, we will focus
on the {\it non-Markovian} aspects of decoherence. Non-Markovian
dynamics, often neglected in the literature (models are mainly
based on a Markov approximation) for technical simplicity, is
actually of crucial importance for the realistic implementations
of quantum information processing. The `effective' model we
consider here invokes a two level simplification from a
multi-level structure. How realistic this is certainly depends on
the way the qubits are defined and realized in the multi-level
structure usually encountered in actual experimental conditions.
Nevertheless our model is able to capture the characteristic short
time behavior common in many physical examples.

Beyond the commonly assumed Ohmic spectrum for the bath, generic
non-Ohmic environments can be studied with this model. Contrary to
the Ohmic case, the sub-Ohmic environment (including
$1/f^{\alpha}$ type) causes nontrivial long time behavior such as
anomalous diffusion or localization \cite{LCDFGZ87} owing to the
long range temporal bath correlation. As demonstrated in
\cite{NPYT88}, the influence of slow environment can be
dynamically decoupled from the system by using relatively slow
pulses\cite{BB}. In the present paper, we will mainly focus on the
opposite case of supra-Ohmic
environments\cite{Weiss99,LCDFGZ87,GargKim89}. Owing to the
ultra-short time bath correlations,  nontrivial short-time system
dynamics enters, which is particularly difficult to describe by
means of other models or approximations. The decoherece time scale
in the supra-Ohmic environment can be much shorter than the one in
the Ohmic case and thus is hard to remove by external pulses. Thus
supra-Ohmic environment can be a major obstacle for the
realization of quantum computation and information processing.

We need to emphasize that to fully  follow the coherence of an
open system where self-consistency is required, because of the
back-action from the environment, we need to study non-Markovian
processs. We will also argue that, for a generic class of
environment, Markovian approximations are not strictly valid. To
facilitate comparison with results in related papers we will
compare our methods with other commonly used approximations to the
spin-boson model, such as the Born approximation and  the
Born-Markov rotating-wave approximation\cite{QOtext} for two-level
and multi-level systems.

The outline of this paper is as follows: In Sec.~\ref{QBME} we
specify the model and cast it in the influence functional
formalism in the presence of an external field. In
Section~\ref{sec:SBQBM} we outline our idea of an effective 2LS
using the QBM approach. Then we make correspondence between the
phase space representation discussed in Sec.~\ref{QBME} with the
Fock space representation. We compare our approach with other
methods based on Born-Markov and rotating-wave-approximation. Our
results are presented in Sec.~\ref{Results}. In
Sec.~\ref{Discussion} we discuss the limitations and potential
extensions of this approach.

\section{QBM in the presence of an external field}
\label{QBME}

\subsection{The model}
Our model consists of a Brownian particle interacting with a
thermal bath in the presence of an external field. We follow the
notion developed in \cite{QBM1,ShiokawaKapral02}. (We use the
units in which $k_B=\hbar=1$.) The Hamiltonian for this model can
be written as
\begin{eqnarray}
   H = H_S +  H_B + H_I + H_F,
      \label{H}
\end{eqnarray}
where the dynamics of the system $S$ (with coordinate $x$ and
momentum $p$) is described by the Hamiltonian
\begin{eqnarray}
    H_S = \frac{ {p}^2}{2M} + V_{0}(x),
      \label{HC}
\end{eqnarray}
and the (bare) potential $V_{0}(x)$ is related to the physical
potential by a counter term $\Delta  {V}$  i.e. $V_{0}(x)=V(
{x})+\Delta  {V}$ (see below). The Hamiltonian of the bath is
assumed to be composed of
 harmonic oscillators with natural frequencies $\omega_{n}$ and masses $m_{n}$,
\begin{eqnarray}
 H_{B} = \sum_{n=1}^{N}( \frac{ {p}_{n}^{2}}{2 m_{n}}
    + \frac{m_{n} \omega_{n}^{2} {q}_{n}^{2}}{2}).
     \label{HB}
\end{eqnarray}
where ($ {q}_1,..., {q}_N, {p}_1,..., {p}_N)$ are the coordinates
and their conjugate momenta.
The interaction between the system S and the bath B is assumed to
be bilinear,
\begin{eqnarray}
  H_{I} =  {x} \sum_{n=1}^{N} c_{n}  {q}_{n},
    \label{HI}
\end{eqnarray}
where $c_n$ is the coupling constant between the Brownian
oscillator and the $n$th bath oscillator with coordinate $q_n$.
The coupling constants are related to the spectral density
$J_B(\omega)$ of the bath by,
\begin{eqnarray}
J_B(\omega) \equiv \pi \sum_{n}\frac{c_{n}^2}{2 m_n \omega_n}
\delta(\omega - \omega_n). \label{SpectralDensity}
\end{eqnarray}
We asssume the spectral density has the form
\begin{equation}
J_{B}(\omega) = 2 M \gamma \omega^{\nu} e^{-\omega/\Lambda},
\end{equation}
where $\nu=1$ is Ohmic, $\nu<1$ is sub-Ohmic , and $\nu>1$ is
supra-Ohmic. We will discuss the Ohmic and supra-Ohmic ($\nu=3$)
cases in detail.

The counter term $\Delta V$ depends on $c_n$, $m_n$, $\omega_n$,
$p$ and $x$ and is given by
\begin{eqnarray}
 \Delta V =
\left\{
\begin{array}{ccc}
2 M \gamma \Lambda x^2 / \pi
&\hspace{1cm} & (\nu=1) \\
2 M \gamma \Lambda p^2 / M^2 \pi + 2 M \gamma \Lambda^3 x^2 / \pi
&\hspace{1cm} &(\nu=3).
\end{array}
\right. \label{counterterm}
\end{eqnarray}
This term is introduced to cancel the shift in the mass and
frequency of the Brownian oscillator due to its interaction with
the bath which will become divergent when the frequency cutoff
$\Lambda \rightarrow \infty$. As is customary, we consider the
renormalized quantities after including a counter term as the
physical observables with specified values.


For a linear QBM, the potential is
\begin{eqnarray}
  V(x) = \frac{M \Omega^2 x^2}{2},
    \label{HI}
\end{eqnarray}
where $\Omega$ is the natural frequency of the system oscillator
%

Finally, the Hamiltonian for the external field is
\begin{eqnarray}
  H_{F} =  -x E(t),
    \label{HI}
\end{eqnarray}
where $E(t)$ is the external field.

\subsection{The Influence Functional}
\label{app:Formulas}

In this subsection, we make connection with the treatment of QBM
based on the influence functional\cite{FeyVer63} with a phase
space representation for the Wigner function\cite{Wigner32}. First
we consider the case without an environment. We define the
transition element between the initial state $| x_0 ~q_{0}
\rangle$ at $t=0$ and the final state $| x ~q \rangle$ at time $t$
to be
\begin{eqnarray}
        {K}(x,q;t \mid x_0, q_{0};0)
 \equiv \langle x ~q | e^{- i H t} | x_0 ~q_{0} \rangle.
          \label{KernelK}
\end{eqnarray}
The Liouville equation for the density matrix is
\begin{eqnarray}
i \frac{\partial}{\partial t}  {\rho}(t) = \left[ H,  {\rho}(t)
\right], \label{Liouville}
\end{eqnarray}
where $\left[ ,  \right]$ is the commutator. In the coordinate
representation, the density matrix becomes
\begin{eqnarray}
 {\rho}(x,x',q,q',t)
    \equiv \langle x ~q |  {\rho}(t) | x' ~q' \rangle
\end{eqnarray}
with the collective notation for bath variables $q \equiv \{ q_{n}
\}$. The time evolution of the density matrix is given by
\begin{eqnarray}
   {\rho}(x,x',q,q',t) =
       \int d x_{0} d x'_{0} d q_{0} d q'_{0}
        {K}(x,q; t \mid x_0, q_{0};0)
   {\rho}(x_0, x'_0, q_{0}, q'_{0}, 0)
    {K}^{*}(x',q'; t \mid x'_0, q'_{0};0) ,
          \label{DMintegral}
\end{eqnarray}

In the present paper, we assume that the characteristic time scale
for the bath is much shorter than the system. Under this
condition, we may integrate out the bath harmonic oscillator
variables to obtain an equation for the reduced density matrix
${\rho}_r(x,x') \equiv \int dq {\rho}(x,x',q,q,t)$. For a
factorized initial condition between the system and the bath,
which is assumed to be initially in thermal equilibrium,
\begin{eqnarray}
   {\rho}(x_0, x'_0, q_{0}, q'_{0}, 0) =
   {\rho}_{S}(x_0, x'_0, 0)
  \otimes
  \rho_{B}(q_{0}, q'_{0},0),
          \label{IDMC}
\end{eqnarray}
we can express the time evolution for the reduced density matrix
in an integral form,
\begin{eqnarray}
   {\rho_r}(x,x',t) =
       \int d x_{0} d x'_{0}
        {J}_r(x, x';t \mid x_0, x'_0; 0)
        {\rho}_{S}(x_0, x'_0, 0),
            \label{DMredintegral}
\end{eqnarray}
where its time evolution operator is given by
\begin{eqnarray}
  {J}_r(x, x';t \mid x_0, x'_0; 0) =
  \int d q d q_{0} d q'_{0}
   {K}(x,q;t \mid x_0, q_{0};0)
   \rho_{B}(q_{0}, q'_{0} ,0)
    {K}^{*}(x',q; t \mid x'_0, q'_{0};0)
   \label{Jr}
\end{eqnarray}
For a harmonic oscillator bath, we have the exact expression
\begin{eqnarray}
J_{r}(x, x';t \mid x_0, x'_0; 0)
  \equiv \int^{(x x')}_{(x_0 x'_0)} {\mathcal D} x {\mathcal D} x'
  e^{i {\cal S}[x,x']}.
\nonumber \\
\label{JF}
\end{eqnarray}
The total action ${\cal S}[x,x']$ consists of several
contributions:
\begin{eqnarray}
{\cal S}[x,x']&=& {\cal S}_S[x,x'] + \Delta {\cal S}_C[x,x']+{\cal
S}_{E}[x,x'] +{\cal S}_{IF}[x,x'],
\end{eqnarray}
where the sum of the actions for the system $S$ plus its counter
action is given by
\begin{eqnarray}
&&({\cal S}_{S}+\Delta {\cal S}_S)[R,r]= \int_{0}^{t} ds \{ M_{0}
\dot{R}(s)\dot{r}(s) - M_{0}\Omega_{0}^2 R(s)r(s) \},
\end{eqnarray}
where $R \equiv (x+x')/2$, $r \equiv x-x'$ and for notational
convenience, we have assumed the bare mass $M_{0}$ and bare
frequency $\Omega_{0}$ take on the values $M_{0}=M$ and
$M_{0}\Omega_{0}^2=M\Omega^2+ 4 M \gamma \Lambda/\pi$ for $\nu=1$
while $M_{0}=M+4 M \gamma \Lambda/\pi$ and
$M_{0}\Omega_{0}^2=M\Omega^2+ 4 M \gamma \Lambda^3/(3\pi)$ for
$\nu=3$. The action for the external field is
\begin{eqnarray}
{\cal S}_{E}[R,r] &=& \int_{0}^{t} ds r(s) E(s). \label{SE}
\end{eqnarray}
The influence action ${\cal S}_{IF}[x,x']$ accounts for the effect
of the bath on $S$ and is given by
\begin{eqnarray}
&&{\cal S}_{IF}[R,r] = i \int_{0}^{t} ds \int_{0}^{s} ds' r(s)
\mu(s-s') r(s') \nonumber \\
&&\qquad \qquad - 2 \int_{0}^{t} ds \int_{0}^{s} ds' r(s)
\nu(s-s') R(s'), \label{SAC2}
\end{eqnarray}
where
\begin{eqnarray}
\nu(t)&=& \frac{1}{\pi} \int_{0}^{\infty} d \omega J_B(\omega)
\coth \frac{\beta \hbar \omega}{2} \cos \omega t, \\
\mu(t)&=&-\frac{1}{\pi} \int_{0}^{\infty} d \omega J_B(\omega)
\sin \omega t \label{noisekernel}
\end{eqnarray}
are the noise and dissipation kernels respectively.

From Eqs.~(\ref{SAC2}) the Euler-Lagrange equations for $R$ and
$r$ are
\begin{eqnarray}
 M_{0} \ddot{R}_c(t)
+  M_{0} \Omega_{0}^2 R_c(t) &+& 2 \int_{0}^{t} ds \mu(t-s)
R_c(s)=E(t), \label{EL1}
\end{eqnarray}
\begin{eqnarray}
M_{0} \ddot{r}_c(s) +
 M_{0} \Omega_{0}^2 r_c(s)
&-& 2 \int_{s}^{t} ds' \mu(s-s') r_c(s')=0. \label{EL2}
\end{eqnarray}
These equations have nonlocal kernels which contain the
information of the past history of the bath in the presence of the
system variables. Because of this, these equations normally
contain time derivatives higher than two. As a result, they admit
unphysical solutions. These unphysical solutions are removed by an
order reduction procedure, reducing them into second order
differential equations with well-defined initial value problems.
They can also be specified uniquely by imposing the initial and
final conditions: $R_0$ and $R_t$ ($r_0$ and $r_t$).

If we let the two independent solutions of the homogeneous part of
Eq.~(\ref{EL1}) (Eq.~(\ref{EL2})) be $u_i(s)$($v_i(s)$), $i=1,2$,
with boundary conditions $u_1(0)=1,u_1(t)=0$, $u_2(0)=0,u_2(t)=1$
($v_1(0)=1,v_1(t)=0$, $v_2(0)=0,v_2(t)=1$), the solutions of
these uncoupled equations can be written as
\begin{eqnarray}
R_c(s) &=& R_0 u_1(s)  + R_t u_2(s) + e(s),
\nonumber \\
r_c(s) &=& r_0 v_1(s)  + r_t v_2(s), \label{SolutionRr}
\end{eqnarray}
where $e(t)=\int_{0}^{t} ds g_{+}(t-s)E(s)/M $. The solutions
$v_1$ and $v_2$ satisfy the homogeneous part of the backward time
equation (\ref{EL2}) and are related to $u_1$ and $u_2$ by
$v_1(s)=u_2(t-s)$ and $v_2(s)=u_1(t-s)$. The function $g_{+}(s)$
($g_{-}(s)$) also satisfies the homogeneous part of
Eq.~(\ref{EL1}) (Eq.~(\ref{EL2})) with boundary conditions
$g_{\pm}(0)=0,\dot{g}_{\pm}(0)=1$. The solutions for $g_{\pm}$ for
Ohmic and supra-Ohmic cases are given in Appendix A of
\cite{ShiokawaKapral02}. From these solutions $u_{1,2}$ and
$v_{1,2}$ can be determined.

Since the potentials in our model are harmonic, an exact
evaluation of the path integral can be carried out. It is
dominated by the classical solution given in (\ref{SolutionRr}).
From these classical solutions, we write the action ${\cal
S}[x,x']$ as
\begin{eqnarray}
{\cal S}[R_c,r_c]
&=& \Big( M \dot{u}_1(t) R_0 + M \dot{u}_2(t) R_t\Big) r_t \nonumber \\
&-& \Big( M \dot{u}_1(0) R_0 +  M \dot{u}_2(0) R_t\Big)r_0 \nonumber \\
&+& i \Big( a_{11}(t) r_0^2 + ( a_{12}(t) + a_{21}(t)) r_0 r_t +
a_{22}(t) r_t^2 \Big) \nn &+& e_1(t) r_0 + e_2(t) r_t \;.
\label{SQBM}
\end{eqnarray}
Here $(e_1(t),e_2(t)) = {\bf e}^{T} = \int_{0}^{t} ds
(v_1(s),v_2(s)) e(s)$ and
\begin{eqnarray}
a_{kl}(t) &=& \frac{1}{2} \int_{0}^{t} ds \int_{0}^{t} ds' v_k(s)
\mu(s-s') v_l(s'), \label{aij}
\end{eqnarray}
for $(k,l=1,2)$ contains the effects of induced fluctuations from
the bath on the system dynamics.


Using the results above, $J_r$ in Eq.~(\ref{JF}) can be written in
the compact form,
\begin{eqnarray}
J_r(R_t, r_t;t  \mid R_0, r_0; 0)   =
     N(t) e^{i {\mathcal L} },
  \label{Jrsimple}
\end{eqnarray}
where ${\mathcal L}= {\bf R}^{T} {\bf u} {\bf r} +i {\bf
r}^{T}{\bf a} {\bf r} +{\bf e}^{T} {\bf r}$, $({\bf a})_{ij} =
a_{ij}$, ${\bf R}^T=(R_0, R_t)$ and ${\bf r}^T=(r_0, r_t)$
\begin{eqnarray}
{\bf u} =
        \left( \begin{array}{cc}
       u_{11}  & u_{12}   \\
       u_{21}  & u_{22}
         \end{array}      \right)
 \equiv
       M  \left( \begin{array}{cc}
       - \dot{u}_{1}(0) & \dot{u}_{1}(t)  \\
       - \dot{u}_{2}(0) & \dot{u}_{2}(t)
         \end{array}      \right)\;,
\label{U}
\end{eqnarray}

\subsection{QBM in the phase space representation}
\label{sec:PHASE}

The Wigner function is related to the density matrix by
\begin{eqnarray}
W_{r}(R,P,t)= \frac{1}{2\pi} \int dr e^{-iPr}
\rho_r(R+r/2,R-r/2,t).
\end{eqnarray}
The Wigner distribution function obeys the evolution equation
\begin{eqnarray}
W_{r}(R_t,P_t,t)=
 \int dR_0 dP_0 \; K(R_t, P_t;t \mid R_0, P_0; 0)
 W_{r}(R_0,P_0,0), \label{WrdensityC}
\end{eqnarray}
where $K(R, P;t \mid R_0,P_0;0)$ is defined by
\begin{eqnarray}
K(R, P;t \mid R_0,P_0;0)= \frac{1}{2 \pi} \int dr dr_0\;
e^{-i(Pr-P_0r_0)} J_r(R, r;t  \mid R_0, r_0; 0)\;.
\end{eqnarray}
The propagator $K$ for the Wigner function is given by
\begin{eqnarray}
 K(R, P; t \mid R_0, P_0; 0)
&=& \frac{N(t)}{2 \pi} \int dr dr_0 \;
e^{i (-P r + P_0 r_0+ {\mathcal L})/\hbar} \nonumber \\
 &=& N_W(t)
 \exp\Big[ - \delta \vec{X}^{T}
{\bf \Sigma}^{-1} \delta \vec{X} ] \Big],\nonumber
  \label{Ksigma}
\end{eqnarray}
where $N_W(t)=N(t)/2 \sqrt{|{\bf a}|}$ and $|{\bf a}|$ is the
determinant of ${\bf a}$. The vector $\delta \vec{X}= \vec{X}
-\langle \vec{X} \rangle$, with
\begin{eqnarray}
\vec{X} = \left( \begin{array}{c}
        R \\
        P-e_2
       \end{array}      \right) \;,
\label{RWvec}
\end{eqnarray}
and
\begin{eqnarray}
\langle \vec{X} \rangle= \left( \begin{array}{c}
        \langle R \rangle \\
        \langle P \rangle
         \end{array}      \right)
=  \left( \begin{array}{cc}
       C_{11}  & C_{21}\\
       C_{12}  & C_{22}
         \end{array}      \right)
\left( \begin{array}{c}
        R_0 \\
        P_0+e_1
         \end{array}      \right)
=\frac{-1}{u_{21}}
   \left( \begin{array}{cc}
       u_{11} & 1 \\
       |{\bf u}|  & u_{22}
         \end{array}      \right)
\left( \begin{array}{c}
        R_0 \\
        P_0+e_1
         \end{array}      \right).
\label{RC}
\end{eqnarray}
Here ${\bf \Sigma}$ is a matrix characterizing the induced
fluctuations from the environment:
\begin{eqnarray}
&&{\bf \Sigma} = \frac{2}{u_{21}^2}
       \left( \begin{array}{cc}
        a_{11}                       & a_{12} u_{21} - a_{11} u_{22} \\
       a_{12} u_{21} - a_{11} u_{22} &
       a_{11} u_{22}^2 - 2 a_{12} u_{21} u_{22} + a_{22} u_{21}^2
         \end{array}      \right).
\label{QW2}
\end{eqnarray}
At long times, fluctuations of the system are goverened by these
terms as $\Sigma_{11} \rightarrow \langle (\Delta R)^2 \rangle$,
$\Sigma_{22} \rightarrow \langle (\Delta P)^2 \rangle$,
$\Sigma_{12}=\Sigma_{21} \rightarrow 0$.

It is seen that this solution for the density matrix obeys
non-Markovian dynamics in that the solution at a given time
depends on its past history. Owing to the time dependent nature of
their coefficients,  despite its simple appearance, these
equations are not easy to solve without approximations. The
commonly used Markovian approximations may miss the essential
features in the description of quantum/classical correspondence:
it tends to underestimate the loss of quantum coherence because
the rapid initial increase of diffusion coefficients is crucial
for decoherence at low temperature in the strong coupling case. It
is simply not a good approximation for a harmonic oscillator model
with a generic spectral density.


\section{Effective spin-boson model from QBM}
\label{sec:SBQBM}
\subsection{Dynamical level reduction}

We first illustrate our scheme of dynamical level reduction based
on the harmonic QBM, which can be viewed as an infinite-level
system in a bosonic environment
 \begin{eqnarray}
    H_S^{(\infty)} +H_{I}^{(\infty)} = \Omega a^{\dagger} a + \sqrt{2 \Omega}
    (a + a^{\dagger}) \sum_{n=1}^{N_B} c_{n}
    {q}_{n}
\end{eqnarray}
This can be viewed as a limit of the finite $N$-level system:
\begin{eqnarray}
H_S^{(N)} +H_{I}^{(N)} = \Omega S_{N}^{+} S_{N}^{-} + (S_{N}^{-} +
S_{N}^{+}) \sum_{n=1}^{N_B}
   \tilde{c}_{n}  {q}_{n}
         \label{NLS}
\end{eqnarray}
when $N\rightarrow \infty$. Here we have absorbed $\sqrt{2
\Omega}$ by defining $\tilde{c}_{n}= \sqrt{2 \Omega} c_{n}$.

At finite temperature $T$, only those modes up to $N \sim k_B T /
\hbar \Omega$ are occupied. Thus at low temperature $T \sim \hbar
\Omega$, the effective number of levels of harmonic QBM is
significantly reduced. In particular, at $k_B T < \hbar \Omega$,
we expect that the system is effectively reduced to two-levels:
\begin{eqnarray}
    H_S^{(2)} +H_{I}^{(2)} =\Omega S_{2}^{+} S_{2}^{-} + (S_{2}^{-} + S_{2}^{+})
   \sum_{n=1}^{N_B}  \tilde{c}_{n}  {q}_{n}.
      \label{TLS}
\end{eqnarray}
The formal correspondence is achieved by replacing the harmonic
oscillator annihilation/creation operator $a,a^{\dagger}$ by the
two-level pseudo spin annihilation/creation (Pauli) operator
$S_{2}^{-},S_{2}^{+}$. The spin-boson model can be obtained by
rewriting the Pauli operators as
\begin{eqnarray}
H_S^{(2)} +H_{I}^{(2)} = \Omega (S_{2}^{z}+\frac{1}{2}) +
S_{2}^{x} \sum_{n=1}^{N_E} \tilde{c}_{n} {q}_{n}.
      \label{TLR}
\end{eqnarray}

\subsection{Fock states from phase space representation}
The correspondence between the Fock state representation for the
pseudo spin qubits and the phase space representation is given as
follows. First we write the density matrix in terms of the phase
space variable as
\begin{eqnarray}
   \hat{\rho}(t) &=&
 \int \frac{d^2 z}{\pi}\chi_Q(z,\bar{z},t) e^{- i z a }e^{-i \bar{z}
 a^{\dagger}},
             \label{DMWigner}
\end{eqnarray}
where
\begin{eqnarray}
  \chi_Q(z,\bar{z},t) = {\mbox Tr} \left[ \hat{\rho}(t)e^{i z a }e^{i \bar{z} a^{\dagger} }
            \right] \label{chiQ}
\end{eqnarray}
is a characteristic function for the Q
representation\cite{MandelWolf95,Twamley93}. In a Fock space
representation,
\begin{eqnarray}
 \rho_{kl}(t) &=&
  \int \frac{d^2 z}{\pi}\chi_Q(z,\bar{z},t)
  \langle k |  e^{- i \bar{z} a^{\dagger} } e^{- i z a } | l
  \rangle,
                \label{DMkl}
\end{eqnarray}
$\chi_Q(z,\bar{z})$ is related to the characteristic function for
the Wigner representation $ \chi_W(z,\bar{z})$ by
\begin{eqnarray}
  \chi_Q(z,\bar{z},t) =  e^{- \frac{|z|^2}{2}}
  \chi_W(z,\bar{z},t).
                \label{chiQchiW}
\end{eqnarray}
These characteristic functions are Fourier components of the phase
space distribution functions, thus
\begin{eqnarray}
  \chi_Q(z,\bar{z},t) =  \int d^2 \alpha Q(\alpha,\bar{\alpha})
  e^{i \bar{z}  \bar{\alpha}} e^{i z \alpha }\\
\chi_W(z,\bar{z},t) =  \int d^2 \alpha W(\alpha,\bar{\alpha})
  e^{i \bar{z}  \bar{\alpha}} e^{i z \alpha }.
                \label{chiQchiWQW}
\end{eqnarray}

The characteristic function $\chi_Q(z,\bar{z},t)$ for the harmonic
QBM evolved from the initial ground state has the following
Gaussian form:
\begin{eqnarray}
 \chi_Q(z,\bar{z},t) = \exp\left[
 - a(t) z^2 - \bar{a}(t) \bar{z}^2
 -2b(t)|z|^2
 +i \alpha_f(t) z +i \bar{\alpha_f}(t) \bar{z}\right] \label{chiQgaussian}
\end{eqnarray}
The time dependent coefficients $a\equiv c+\sigma$, $b\equiv
C+\Sigma+1/4$, and $\alpha_f$ in the above have their origins in
the classical trajectory ${\bf C}$ of a damped harmonic oscillator
given in Eq. (\ref{RC}), the induced fluctuations ${\bf \Sigma}$
from the bath given
 in Eq.(\ref{QW2}), and the external field $E$. The relations of these
components are given as follows:
\begin{eqnarray}
8 c&=&  C_{22}^2 -  C_{11}^2 +  \Omega^2 C_{12}^2
-\frac{C_{21}^2}{\Omega^2} + 2 i (C_{11}C_{12}+ C_{21}C_{22}) \nn
8 C&=& C_{11}^2 + \Omega^2 C_{12}^2 + \frac{C_{21}^2}{\Omega^2} +
C_{22}^2 \nn
 4 \sigma&=&    \Omega \Sigma_{11} -
\frac{\Sigma_{22}}{\Omega} +2 i \Sigma_{12} \nn 4 \Sigma&=& \Omega
\Sigma_{11} + \frac{\Sigma_{22}}{\Omega}\label{sS}
\end{eqnarray}
and
\begin{eqnarray}
\alpha_f(t)=\frac{1}{\sqrt{2\Omega}} \int_{0}^{t} \left( \Omega +i
\frac{d}{dt} \right) g_{+}(t-s) E(s), \label{alpha}
\end{eqnarray}
where  $g_{+}$ satisfies the homogeneous part of the equation of
motion in (\ref{EL1}).

From Eq. (\ref{DMkl}) we can directly evaluate the density matrix
in the Fock representation at arbitrary quantum number. For
instance, for an initial ground state, $\hat{\rho}(0)=|0\rangle
\langle 0|$, in the absence of an external field, the ground state
and the first excited state population can be written as
\begin{eqnarray}
   {\rho}_{00}(t) &=&
   \frac{1}{2
\left[ b(t)^2 - |a(t)|^2 \right]^{1/2}
   }
             \label{GSP}
\end{eqnarray}
and
\begin{eqnarray}
   {\rho}_{11}(t) &=&
    \frac{1}{2
\left[ b(t)^2 - |a(t)|^2 \right]^{1/2}
   }   -  \frac{b(t)}{4
\left[ b(t)^2 - |a(t)|^2 \right]^{3/2}
   }.             \label{1SP}
\end{eqnarray}
Let us introduce the Pauli spin representation for the two-level
system:
\begin{eqnarray}
 \langle \sigma_x(t)\rangle &=& {\rho}_{01}(t) + {\rho}_{10}(t)\nn
 \langle \sigma_y(t)\rangle &=& i {\rho}_{10}(t) - i{\rho}_{01}(t)\nn
 \langle \sigma_z(t)\rangle &=& {\rho}_{11}(t) - {\rho}_{00}(t).
             \label{SPRep}
\end{eqnarray}
We can express them by the variables defined in
Eq.(\ref{chiQgaussian})-(\ref{alpha}) for arbitrary two-level spin
initial states as follows:
\begin{eqnarray}
  \langle \sigma_x(t)\rangle &=&
\frac{-1}{ 4 \left[ b(t)^2 - |a(t)|^2 \right]^{3/2}} \left\{
\left[ \langle \sigma_x(0)\rangle C_{22}- \langle
\sigma_y(0)\rangle \Omega C_{12} \right]\left[\mbox{Re}~ a(t)-
b(t)\right] \right. \nn &+& \left. \left[ \langle
\sigma_x(0)\rangle \frac{C_{21}}{\Omega} - \langle
\sigma_y(0)\rangle C_{11}\right] \mbox{Im}~ a(t) \right\},
\label{Sxt}
\end{eqnarray}
\begin{eqnarray}
 \langle \sigma_y(t)\rangle &=&
\frac{-1}{ 4 \left[ b(t)^2 - |a(t)|^2 \right]^{3/2}} \left\{
\left[ \langle \sigma_x(0)\rangle \frac{C_{21}}{\Omega} - \langle
\sigma_y(0)\rangle C_{11}\right] \left[\mbox{Re}~ a(t)+
b(t)\right] \right. \nn &-& \left. \left[ \langle
\sigma_x(0)\rangle C_{22}- \langle \sigma_y(0)\rangle \Omega
C_{12} \right]
 \mbox{Im}~ a(t) \right\},
        \label{Syt}
\end{eqnarray}
and
\begin{eqnarray}
  \langle \sigma_z(t)\rangle &=&
-\frac{b(t)}{ 4 \left[ b(t)^2 - |a(t)|^2 \right]^{3/2}} \\ &+&
\frac{{\rho}_{11}(0)}{ \left[ b(t)^2 - |a(t)|^2 \right]^{3/2}}
\left\{ \mbox{Re}~\left[\bar{c} a(t)\right]- C b(t) - \frac{C}{2}
+\frac{3b(t)}{2} \frac{\mbox{Re}~\left[\bar{c} a(t)\right]- C b(t)
}{ \left[ b(t)^2 - |a(t)|^2 \right]^{3/2}} \right\} \nonumber
\label{Szt}
\end{eqnarray}
The leakage at time $t$ is given by ${\cal L}(t)=1-$min
Tr$\left[\hat{P}\rho_r(t)\right]$, where $\hat{P}$ is the
projection operator onto the computational subspace and the
minimization is taken over initial conditions. In our case,
$\hat{P}=\sum_{n=0,1}| n \rangle \langle n |$. The source of the
leakage in our model is the transition to higher modes. The
leakage is typically estimated by perturbative methods. However,
the exact temporal evolution of this function is highly nontrivial
as we will see below. Note that from the form of our {\it
effective} Hamiltonian in (\ref{TLR}), the behavior of coherence
and population between our model and some others in the literature
(for example, in \cite{LCDFGZ87}) are interchanged. They are
related to each other by a change of basis. We can obtain similar
expressions in the presence of an external field. We will examine
this case in Sec. 4.1. In the Markovian limit, if the limit
exists, two-level spin states become coupled nontrivially and obey
optical-Bloch type equations\cite{API}.

\subsection{Limitations of other approximations}

\subsubsection{Born-Markov approximation}
In this approximation, the bath correlation is neglected. This may
be obtained as a limit of high temperature or slow system
evolution in the Ohmic bath or white noise bath. For a generic
bath spectral density, however, there is no such limit. In a
supra-Ohmic bath, the diffusion constant, when time-averaged for a
long time, vanishes owing to the ultra-short time correlation. In
a sub-Ohmic bath, it diverges owing to the long-range
correlations. Only Ohmic spectrum gives the finite constant
diffusion term.

For weak coupling, the off-resonant counter-rotating terms in the
interaction Hamiltonian are often ignored by invoking the
rotating-wave-approximation (RWA). Although the use of RWA
significantly simplifies the analysis, the dynamics under this
approximation cannot capture the fast dynamics at time scales less
than the natural time scale of the system. Furthermore, the
spectrum of the Hamiltonian under RWA is found to be unbounded
from below\cite{FordOconnell97}. These features suggest that the
range of validity of RWA is restricted to the leading order in the
coupling constant only, where the counter-rotating terms do not
contribute. After neglecting the counter-rotating terms from the
two-level spin-boson Hamiltonian in (\ref{TLR}), we obtain
\begin{eqnarray}
  H_S +H_{SB}
 = \Omega S_{2}^{+} S_{2}^{-} + (S_{2}^{-} + S_{2}^{+}) \sum_{n=1}^{N} c_{n}  {q}_{n}
  \rightarrow
   H_{S} +H_{RWA} = \Omega S_{2}^{+} S_{2}^{-} + \sum_{n=1} c_n ( S_{2}^{-} {b}_{n}^{\dagger}
    + S_{2}^{+} {b}_{n} )
      \label{RWA}
\end{eqnarray}
where ${b}_{n}= (\omega_n q_n +i p_n)/\sqrt{2 \omega_n}$ are bath
annihilation operators. In the presence of an external field,
under the RWA, the reduced density matrix for the Hamiltonian
obeys an optical Bloch equation. This case is commonly described
in quantum optics text books.

\subsubsection{Born-Markov-RWA in multi-level-system (MLS)}
For comparison, we make the same Born-Markov-RWA in our E2L-SBM.
Since the naive high temperature limit of the master equation
obtained from QBM violates positivity\cite{Positivity}, we start
from the master equation in the Lindblad form\cite{QOtext}. For a
particle initially in the Fock state $\hat{\rho}(0)=|k\rangle
\langle k|$, the $Q$ distribution function at time $t$ has the
following form:
\begin{eqnarray}
 Q(\alpha,\bar{\alpha})&=&
  \frac{1}{\pi\left[ 1+ n_B ( 1-e^{-\gamma t})  \right]}
  \exp\left[   -\frac{|\alpha|^2}{1+ n_B ( 1-e^{-\gamma t})}  \right]
   \left[ \frac{(n_B+1)(1-e^{-\gamma t})}{1+n_B ( 1-e^{-\gamma t})}\right]^k \nn
   &\times& \sum_{l=0}^{k} \frac{1}{k!} \frac{k!}{l!
(k-l)!}
    \left[  \frac{|\alpha|^2 e^{-\gamma t}}{(n_B+ 1 )
    ( 1-e^{-\gamma t})\{1+ n_B ( 1-e^{-\gamma t})\}}   \right]^l
    \label{WFock}
\end{eqnarray}
where $n_B\equiv 1/(e^{\beta \Omega}-1)$ is a Planck distribution
factor. For $\hat{\rho}(0)=|1\rangle \langle 1|$, from
(\ref{DMkl}),(\ref{chiQchiW}),and (\ref{WFock}).
\begin{eqnarray}
  \rho_{00}(t) &=&
   \frac{1}{\left[ 1+n_B ( 1-e^{-\gamma t}) \right]^2}
   \left[ 1 -e^{-\gamma t} + n_B ( 1-e^{-\gamma t}) \right]
   \label{DM0011}
\end{eqnarray}
and
\begin{eqnarray}
  \rho_{11}(t) &=&
   \frac{1}{1+n_B ( 1-e^{-\gamma t})}
   \left\{ 1 -  \frac{1+e^{-\gamma t}}{1+n_B ( 1-e^{-\gamma t})}
+\frac{2  e^{-\gamma t} }{\left[ 1+n_B ( 1-e^{-\gamma t})
\right]^2} \right\}
   \label{DM1111}
\end{eqnarray}

\subsubsection{Born approximation}
It is known that any master equation can be written in a
time-convolutionless form\cite{TCL}. However, the exact master
equation in this form is still difficult to deal with. Most
approaches based on the master equation invoke Born-approximation,
then we obtain the tractable form, which can be solved exactly for
simple systems or numerically for others\cite{LossDeVincenzo03}.
The master equation under weak-coupling approximation may be
suitable for describing the short time dynamics but tends to
predict incorrect behavior for long times\cite{Weiss99,LCDFGZ87}.
Our nonperturbative results free from the weak coupling
approximation is applicable to arbitrary time scales.

\begin{figure}[h]
 \begin{center}
\epsfxsize=.45\textwidth \epsfbox{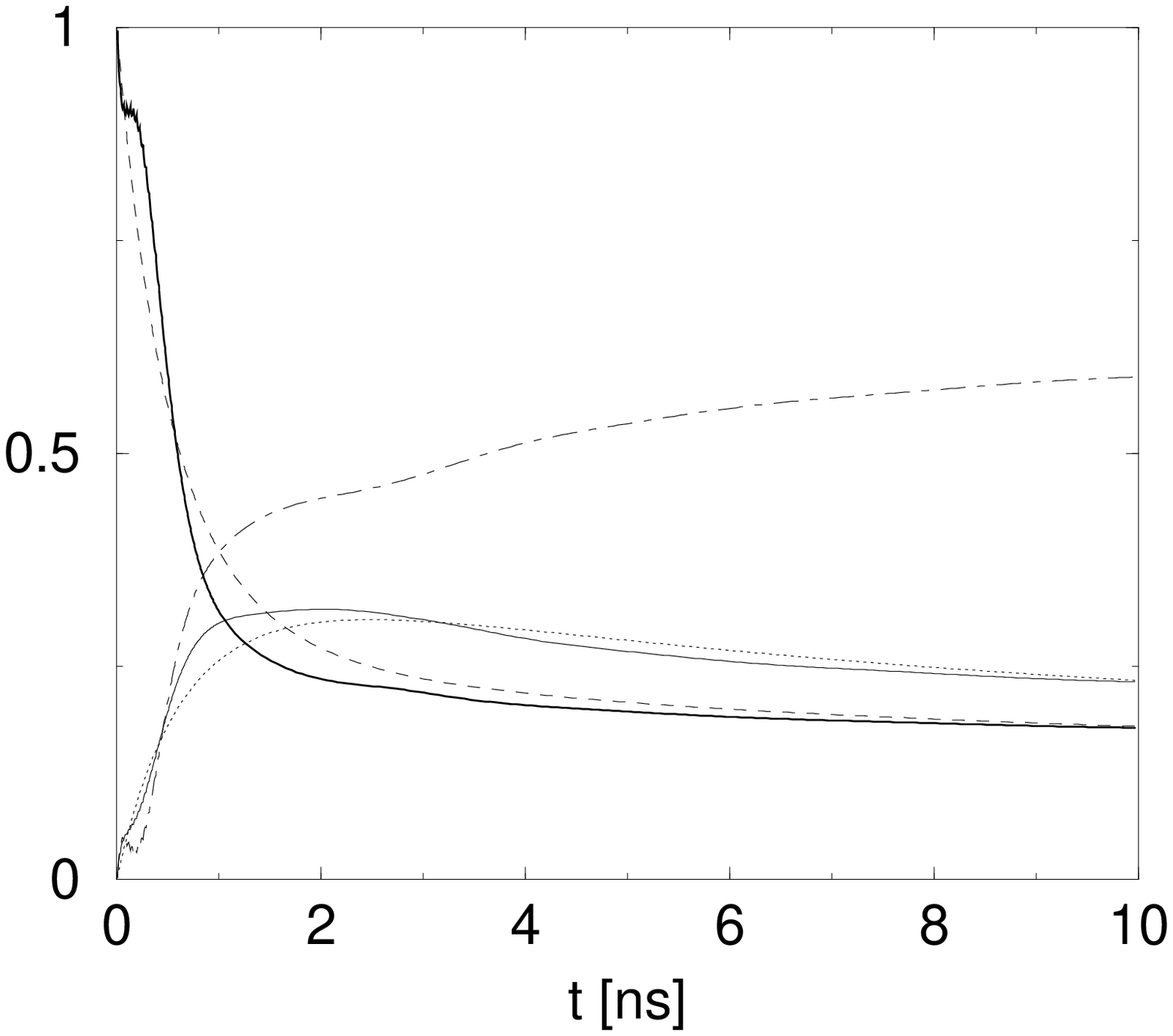}
\epsfxsize=.45\textwidth \epsfbox{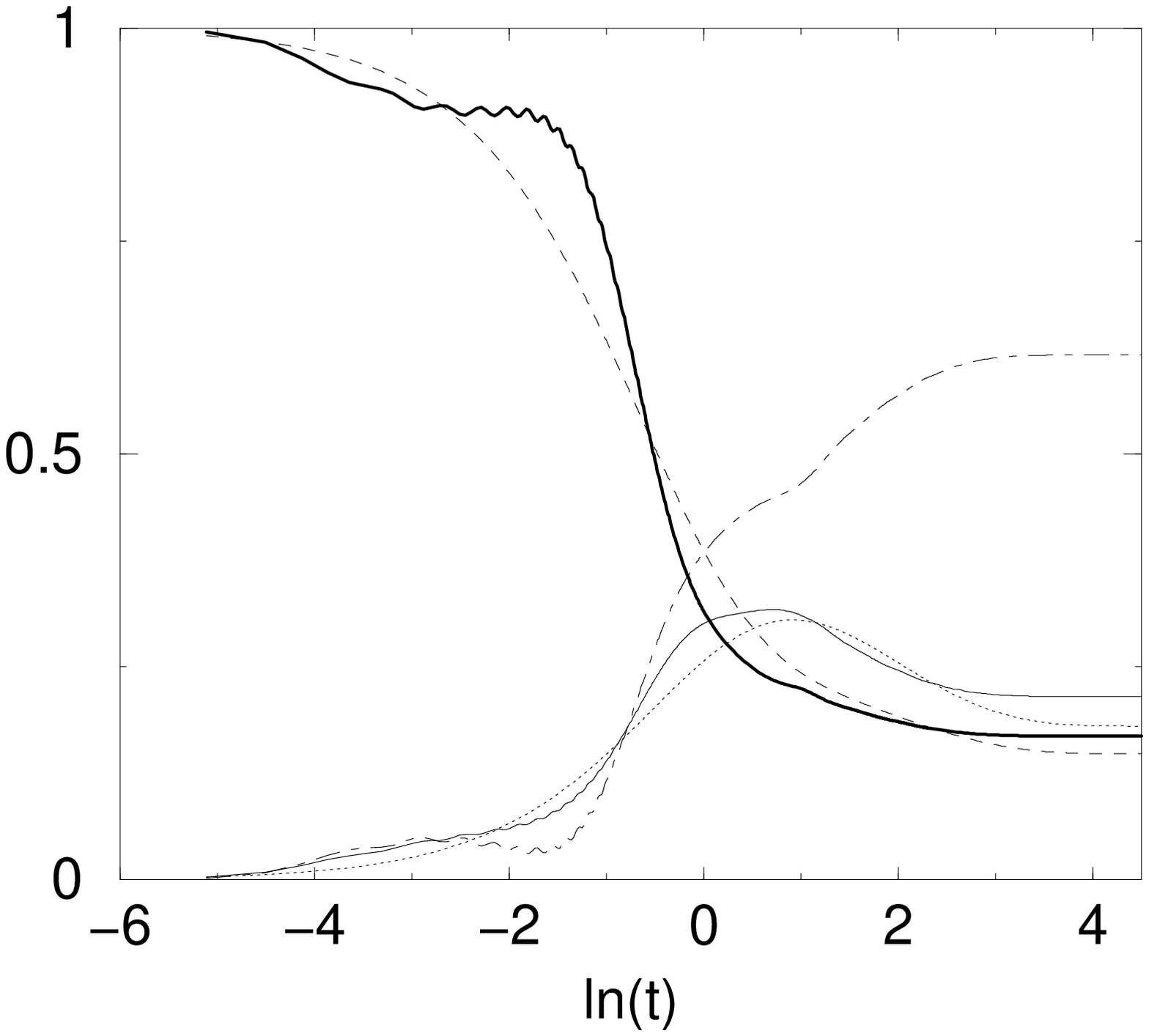}
\end{center}
\caption{ (a) Plot of the time-evolution of the population of the
ground and the first excited state population, and the leakage
(the dot-dashed curve) at $T=50[mK]$ with $\Omega=1[GHz]$,
$\gamma=0.1[GHz]$, $\Lambda=100[GHz]$. The thick solid (dashed)
curve is the exact (Markovian) result for excited states while the
thin solid (dashed) curve is the exact (Markovian) result for
ground states.  Panel (b) is in the logarithmic time
scale.\label{fig1}}
\end{figure}

\section{Results and Discussions}

\subsection{Results} \label{Results}
In Fig. 1, the populations and the leakage at $T=50[mK]$,
$\Omega=1[GHz]$, $\gamma=0.1[GHz]$, and $\Lambda=100[GHz]$ are
shown. The initial state is assumed to be the first excited state.
At this temperature, the exact and the Markovian results agree at
an intermediate time scale (around $t=10[ns]$) but disagree at
initial times. The slow oscillations in Fig. 1b of the exact curve
are from effects due to counter-rotating terms, while the fast
oscillations are due to the frequency cutoff. The large leakage
indicates that at this temperature, $k_B T > \hbar \Omega$, 2LS
description is not a good picture. In Fig. 2, $T=0$ case is shown.
There is a drastic difference in the entire time range shown in
the figure. The exact result follows the quick decay at early
times up to $t\sim10-50[ps]$. Late time decay rate asymptotically
approaches the value given by the Markov approximation. The
leakage is relatively large initially but negligibly small at late
times. This indicates that only the lowest two levels are
essentially populated except for the initial times, $t<5[ns]$. The
initial rapid decay of population is originated in the large
initial leakage due to the transition to noncomputational
subspace. This may be related to the initial large increment of
the diffusion constant in the exact master equation at low
temperature\cite{QBM1}. The total decay slows down as the leakage
is suppressed at an intermediate time scale.
In Fig. 3, the result for a
supra-Ohmic environment is plotted. Compared to the Ohmic case,
the initial decay of the excited state population is much more
drastic but it appears to saturate at late times. Thus if the
initial decay is strong enough, the coherence in the system can be
totally washed out at an early stage, a serious concern for the
quantum devices. On the other hand, if it is small, the system can
remain coherent for a long time. Note that our result disagrees
markedly with the Markovian prediction over the entire time range.

 In Fig. 4, the Rabi oscillations in the presence of an external
sinusoidal pulse at the resonant frequency are plotted. The most
notable difference between the exact results from the Markovian
results is that the exact results show the low onset and low
visibility for all times. The difference is more evident for the
supra-Ohmic case. Our figures also suggest that it is not easy to
determine the characteristics of the environment only from the
experimental Rabi oscillation data without the precise knowledge
of the dissipation. The large increase of leakage is due to the
resonant transition to high level states. Though increasing
anharmonicity in the potential will suppress these transitions to
some extent, the initial rapid increase of leakage is unavoidable
due to energy-time uncertainty relation. In the presence of
tunneling with a biased potential, due to the existence of
resonant transitions to the continuum modes, we expect the result
will be qualitatively similar to our case. In this case, the
leakage in our model can be interpreted as the effect of
tunneling, or more appropriately, environment-induced hopping to
other metastable states.
\begin{figure}[h]
 \begin{center}
\epsfxsize=.45\textwidth \epsfbox{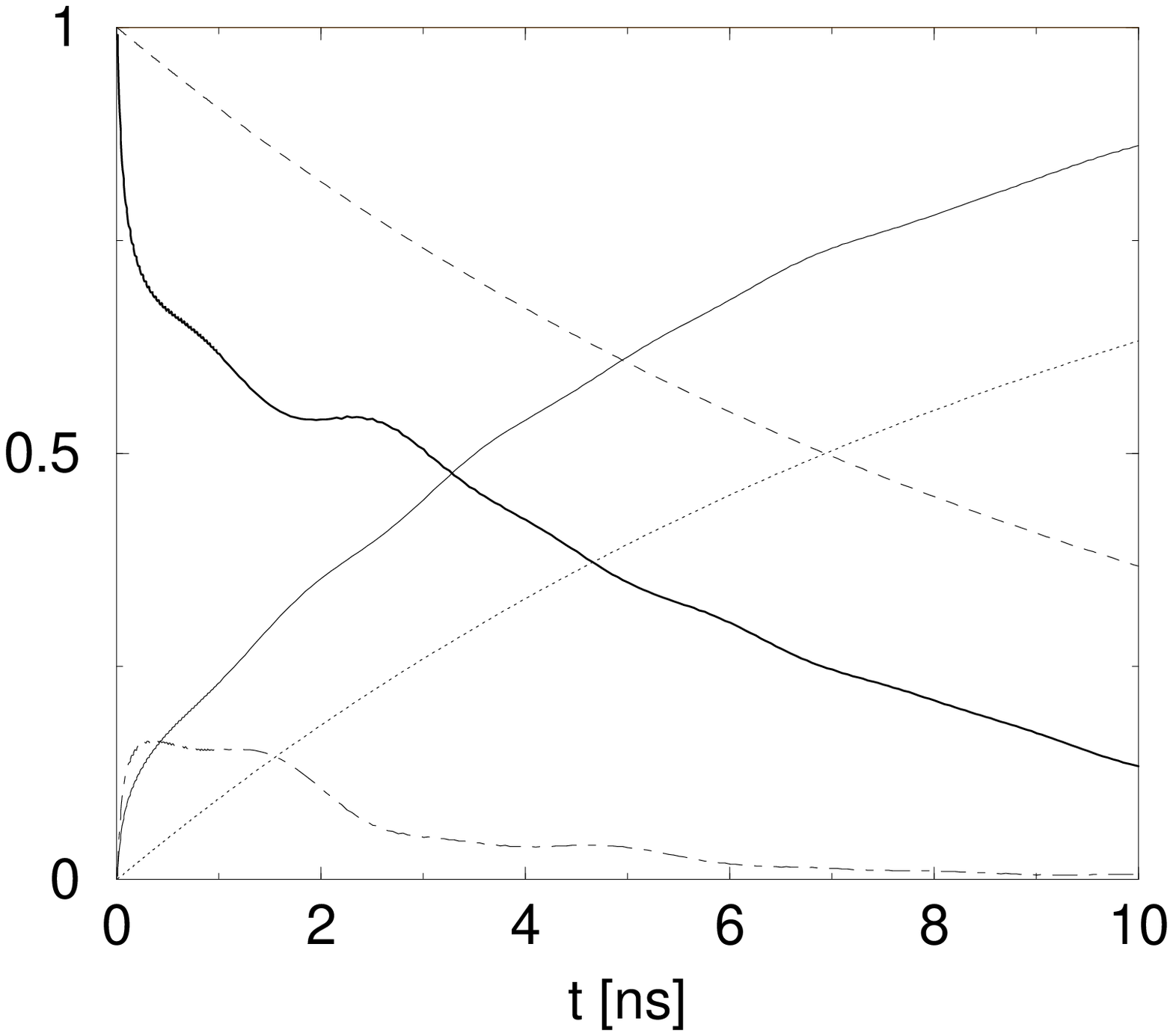}
\epsfxsize=.45\textwidth \epsfbox{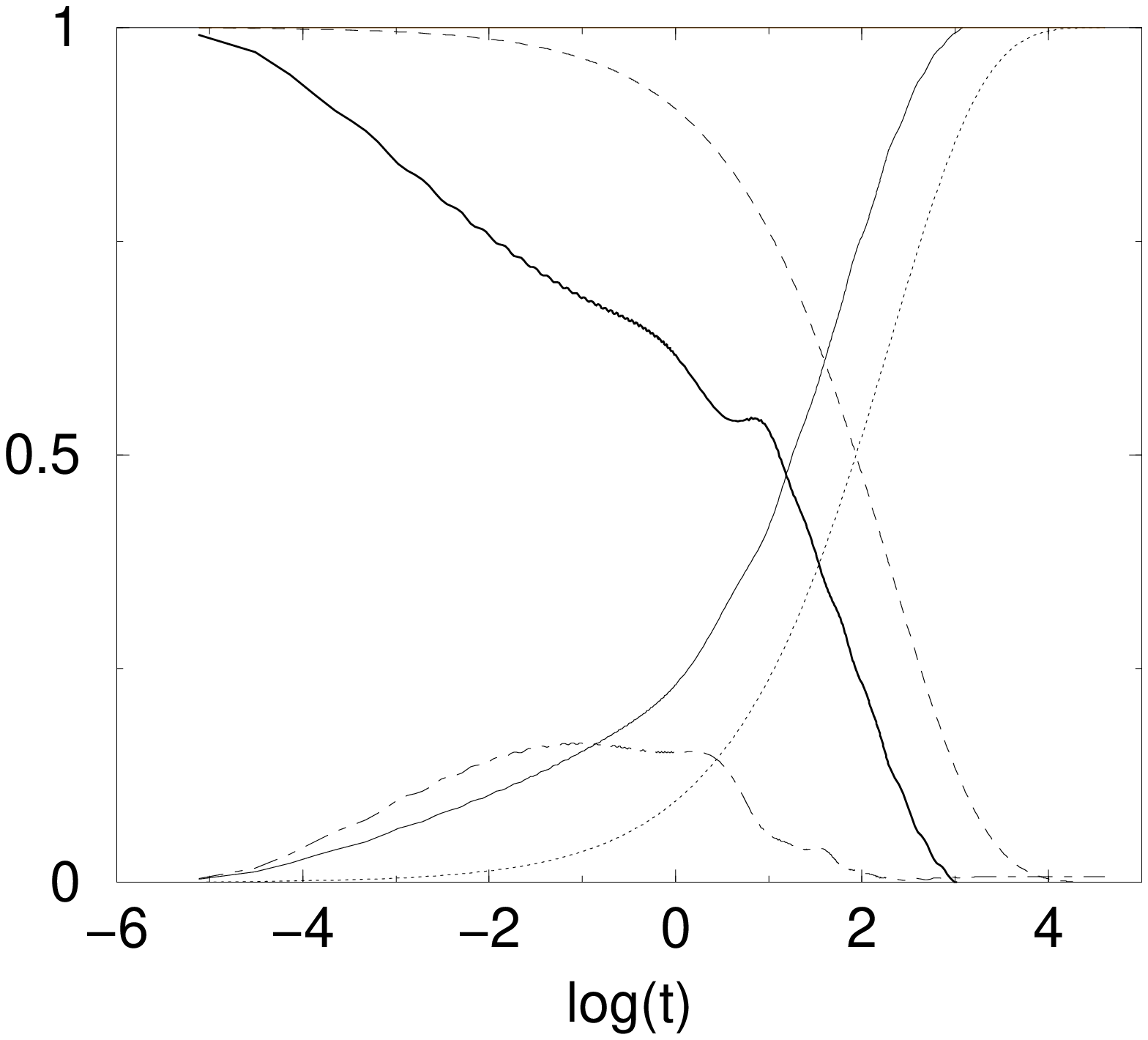}
  \end{center}
  \caption{
The ground and the first excited state population, and the leakage
(the dot-dashed curve) at $T=0$. $\Omega=1[GHz]$,
$\gamma=0.1[GHz]$, $\Lambda=100[GHz]$. The thick solid (dashed)
curve is the exact (Markovian) result for excited states while the
thin solid (dotted) curve is the exact (Markovian) result for
ground states. Panel (b) is in the logarithmic time scale.
\label{fig2}}
\end{figure}
\begin{figure}[h]
 \begin{center}
\epsfxsize=.45\textwidth \epsfbox{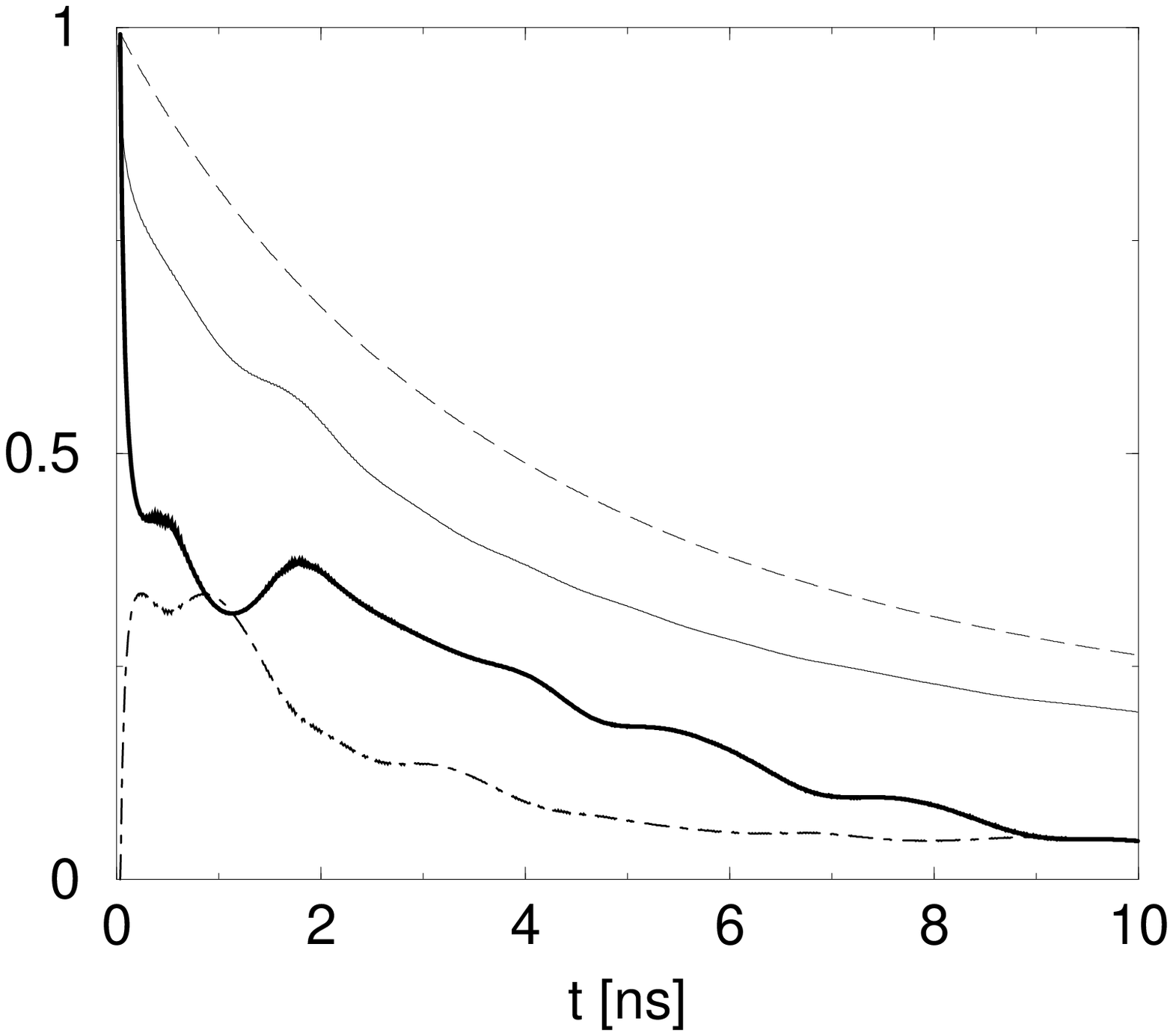}
\epsfxsize=.45\textwidth \epsfbox{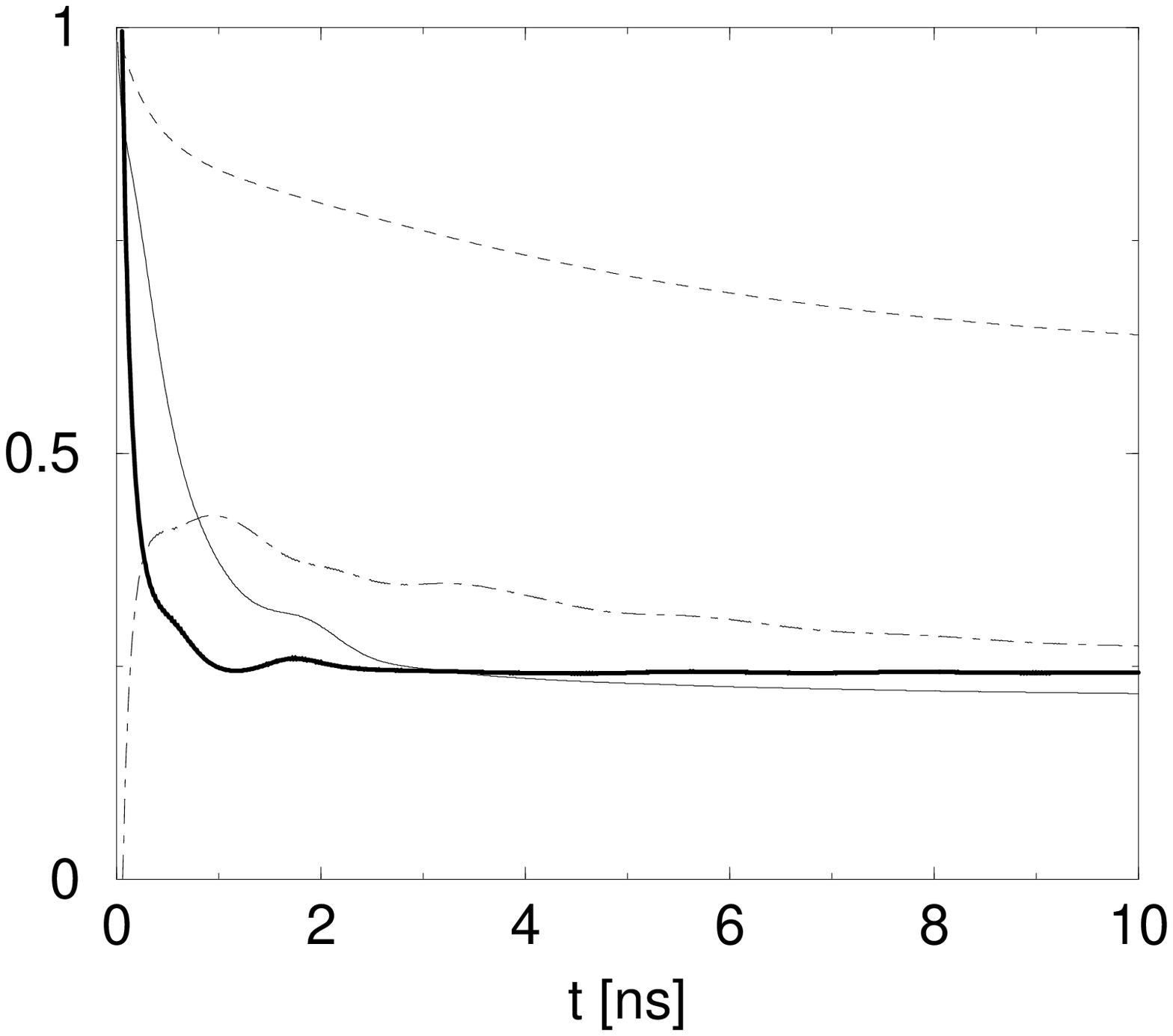}
  \end{center}
\caption{ The excited state populations and the leakage are
plotted for supra-Ohmic environment. $T=10[mK]$ in panel (a) and
$T=50[mK]$ in panel (b). The thick (thin) solid curves is an exact
result for supra-Ohmic (Ohmic) case, while the dashed line is from
the Born-Markov approximation. $\Omega=1.5[GHz]$, $\gamma=0.1$,
$\Lambda=100[GHz]$.  \label{fig5}}
\end{figure}

\begin{figure}[h]
\begin{center}
\epsfxsize=.45\textwidth \epsfbox{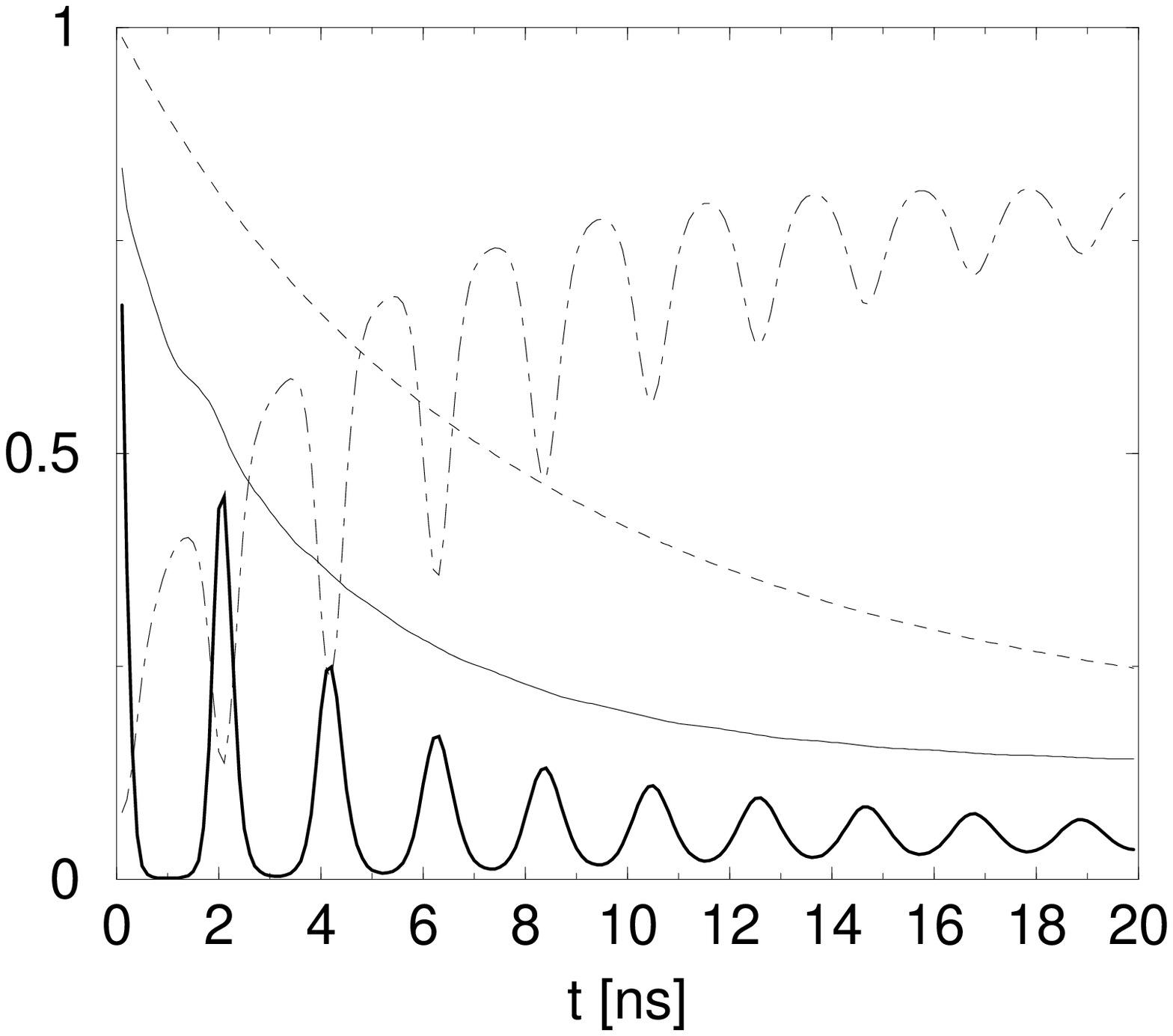}
\epsfxsize=.45\textwidth \epsfbox{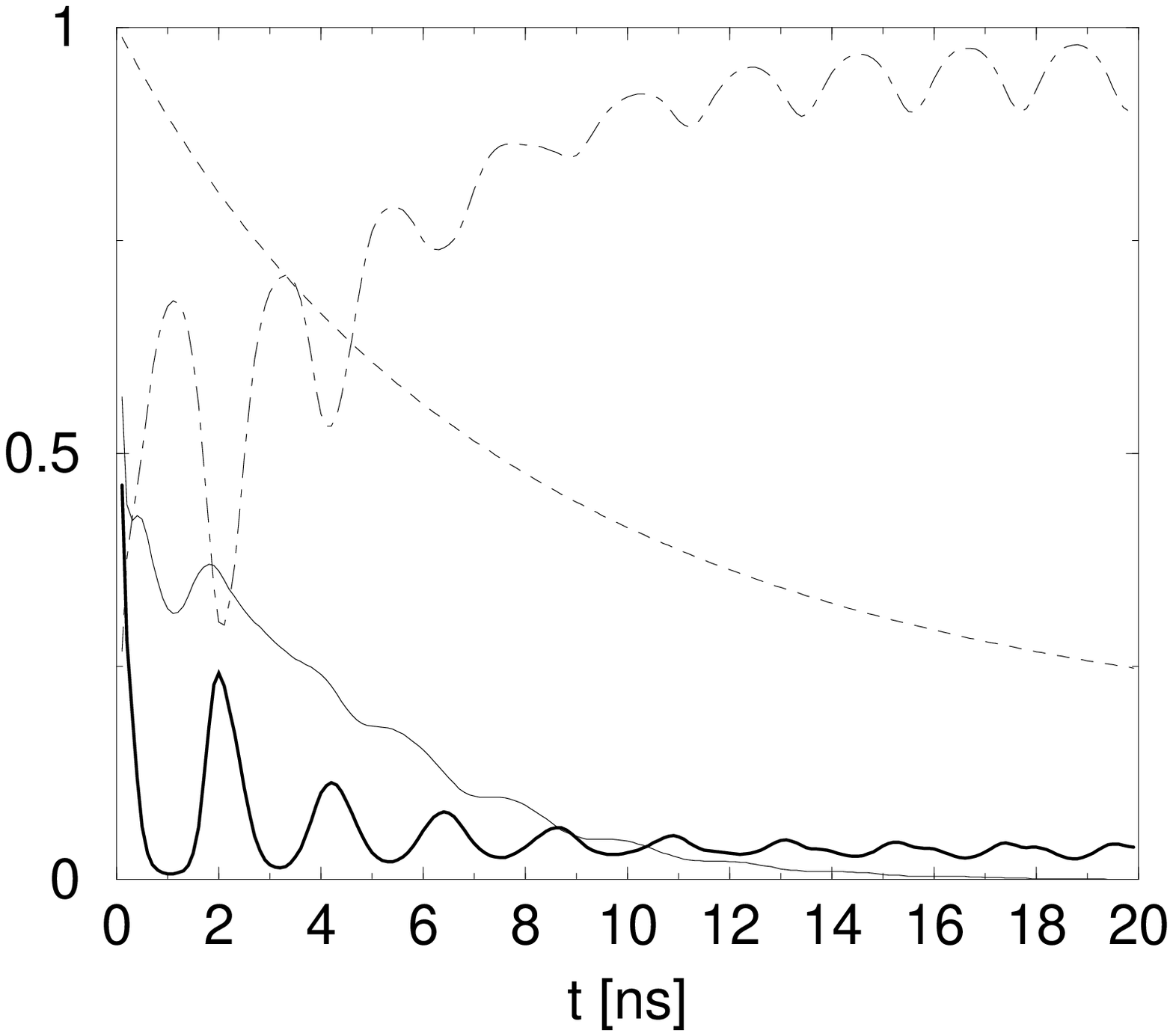}
  \end{center}
\caption{ The excited state populations in the presence and
absence of external field at $T=10[mK]$ are plotted for Ohmic
environment in panel (a) and for supra-Ohmic environment in panel
(b). The applied external is sinusoidal in the form: $E=E_0
\cos(\Omega t)$ with $E_0=1.0$. The thick (thin) curves are exact
results in the presence (absence) of external field, while the
dashed curves are from the Born-Markov apporoximation.
$\Omega=1.5[GHz]$,$\gamma=0.1$,$\Lambda=100[GHz]$. \label{fig6}}
\end{figure}
\newpage

\subsection{Discussion}\label{Discussion}

We saw that at low enough temperatures, many conventional
approaches based on the Born-Markov approximation can
significantly underestimate the environment-induced decoherence
beyond the weak system-bath coupling. In this regime, the
visibility in Rabi oscillations in the exact calculation tends to
be lower than what is expected in the Markovian approximation. Low
visibility in Rabi oscillations is commonly observed in
superconducting qubits\cite{YHCCQW02,MNA02,VACJPUED02,CNHM03}. The
bath time scale is also important in causing the initial rapid
decoherence and leakage; this is completely neglected in analysis
based on the Born-Markov-RWA. This initial effect can manifest
itself as an onset value of Rabi-oscillations. In many practical
implementations of qubits, the temperature of the environment
compared to the bath is small, $k_B T << \hbar \Lambda$, thus we
are in the low temperature regime.

The E2L-SBM approach gives a precise evaluation of the leakage due
to the system's interaction with the environment and the external
control field. For temperatures higher than the characteristic
energy of the oscillator, the large leakage makes the qubit based
on the choice of the lowest two levels ill-defined. During gate
operations, this can become a serious problem and remedies for
stabilizing the system such as using external pulse control may be
necessary. Our result shows that the time scale associated with
leakage is characterized by the dynamical time scale of the system
and the bath.

In realistic macro- or mesoscopic systems, the potential contains
anharmonicity, which causes the deviation of the system dynamics
from the harmonic motion. A measure of anharmonicity near ground
states can be given by the ratio of energy level separation
between the lowest levels $\omega_{01} \equiv \omega_1-\omega_0$
and the excited levels $\omega_{12} \equiv \omega_2-\omega_1$.
When this ratio is small, $\omega_{12}/\omega_{01}<<\omega_{01}$,
the initial short time evolution around a metastable state can be
well-described by the linear dynamics. When the correction to the
energy level due to anharmonicity in the potential becomes
important, it is necessary to include such an effect in our
scheme. Although the large anharmonicity also prevents the leakage
in the long term, the initial large leakage we saw cannot be
completely eliminated for the reason we mentioned before. When we
apply our formalism to the metastable state, eventually the system
state will leave the harmonic oscillator phase space into other
metastable states via tunneling. The harmonic approximation of
coherent dynamics is expected to be accurate at initial times when
the time scales associated with these nonlinear effects are large
compared to the decoherence time scale. In our example, the
deviation from the Markovian prediction is evident in the very
early stage of the system evolution up to $t\sim 1[ns]$ even for
an intermediate temperature. For the realistic implementation of
qubits, the underlying potential landscape leading to the discrete
energy level is already known by
design\cite{YHCCQW02,MNA02,VACJPUED02,CNHM03,JJ,BXRGSJADLW03}. Our
approach based on E2L-SBM is suitable in this situation and will
give a more precise estimate of the open system dynamics than the
one based on the conventional 2LS approximation.
In particular, our results are directly relevant to the
superconducting phase qubit models
\cite{YHCCQW02,MNA02,BXRGSJADLW03}.  In the superconducting
qubits, the major source of decoherence is the noise induced by
the interaction with the current or charge sources mainly during
the qubit manipulations. We have not considered other possible
sources of decoherence such as the coupling to defects or nuclear
and magnetic spins. Multilevel structure in the superconducting
flux qubits was studied in \cite{BKD04} by Born-Markov
approximation without control fields.

For the system-environment coupling we considered in
Eq.(\ref{HI}), the Fock state is not an eigenstate of the
interaction Hamiltonian and is subjected to a complex decay even
under the Born-Markov approximation as shown in Sec. 3.3.2.
Previous study in the high temperature limit indicates the pointer
state under this system-environment coupling is a coherent
state\cite{ZHP93}. Our calculation based on the exact solution for
QBM indicates that, beyond the weak coupling regime, the
environment-induced effect has a crucial impact on the system
dynamics at an early stage.  A factorized initial condition is
used to derive our main results in accord with the initialization
scheme used commonly in quantum information processing
\cite{NielsenChuang}. From the decoherence study of QBM in the
presence of the initial system-bath correlation due to the
preparation effect\cite{Weiss99,RomeroPaz97}, we expect that our
results are robust and should hold for a more general class of
initial conditions.

\section*{Acknowledgments}
This work is supported in part by an ARDA contract.




\begin{thebibliography}{99}

\bibitem{Dec96} {\it Decoherence and the Appearance of the Classical
World in Quantum Theory}, eds. D. Giulini, {\it et al.},
(Springer, Berlin, 1996).

\bibitem{PazZurek00}
J. P. Paz and W. H. Zurek, in {\it Coherent Matter Waves}, Les
Houches Lectures Session LII, (North Holland, Amsterdam, 1999).

\bibitem{Unruh95}
W.G. Unruh, Phys. Rev. A51,992 (1995).


\bibitem{Weiss99} U. Weiss, {\it Quantum Dissipative Systems},
(World Scientific, Sigapore, 1999).

\bibitem{QBM0}
A. O. Caldeira and A. J. Leggett,Physica {\bf A121}, 587 (1983);
V. Hakim and V. Ambegaokar, Phys. Rev. A32, 423 (1985); F. Haake
and R. Reibold, Phys. Rev. A32, 2462 (1985); H. Grabert, P.Schramn
and G. L. Ingold, Phys. Rep. {\bf 168}, 115 (1988); W. G. Unruh
and W. H. Zurek, Phys. Rev. D40, 1071 (1989);
 B. L. Hu and Y. Zhang, Mod.\ Phys.\ Lett.\ A8, 3575 (1993); Int.\
J.\ Mod.\ Phys.\ {\bf A10} (1995) 4537; J. J. Halliwell and A.
Zoupas, Phys. Rev. D52, 7294 (1995); C. Anastopoulos and J. J.
Halliwell, Phys. Rev. D51, 6870 (1995).

\bibitem{QBM1} B. L. Hu, J. P. Paz and Y. Zhang, Phys. Rev. D45,
2843 (1992).

\bibitem{ZHP93}
W. H. Zurek, S. Habib, and J. P. Paz, Phys. Rev. Lett. {\bf 70},
1187 (1993).


\bibitem{LCDFGZ87}
A.~J. Leggett, S. Chakravarty, A.~T. Dorsey, M.~P. A. Fisher, A.
Garg, and W. Zwerger, Rev. Mod. Phys. {\bf 59}, 1 (1987).

\bibitem{TianLloyd00}
L. Tian and S. Lloyd, Phys. Rev. A {\bf 62}, 050301 (2000).

\bibitem{NPYT88}
Y. Nakamura, Yu. A. Pashkin, T. Yamamoto, and J.S.
  Tsai, Phys. Rev. Lett.{\bf 88},
047901 (2002).

\bibitem{BB}
K. Shiokawa and D. A. Lidar, Phys. Rev. A {\bf 69}, 030302 (2004);
H. Gutmann, F. K. Wilhelm, W. M. Kaminsky, and S. Lloyd,
cond-mat/0308107; L. Faoro and L. Viola, quant-ph/0312159; G.
Falci, A. D'Arrigo, A. Mastellone, and E.
Paladino,cond-mat/0312442.

\bibitem{QOtext}
W. H. Luiselle, {\it Quantum Statistical Properties of Radiation},
(Wiley, New York, 1990); H. J. Carmicheal,{\it Statistical Methods
in Quantum Optics},  (Springer, Berlin, 1999).

\bibitem{ShiokawaKapral02}
K. Shiokawa and R. Kapral, J. Chem. Phys. {\bf 117}, 7852 (2002).


\bibitem{FeyVer63}
 R. P. Feynman and F. L. Vernon, Ann. Phys. {\bf 24}, 118 (1963).

\bibitem{Wigner32} E.~Wigner, Phys. Rev. {\bf 40}, 749 (1932).


\bibitem{MandelWolf95}
L. Mandel and E. Wolf, {\it Optical Coherence and Quantum Optics},
(Cambridge University Press,Cambridge,1995).

\bibitem{Twamley93} J. Twamley, Phys. Rev. D48, 5730 (1993).

\bibitem{API}
C. Cohen-Tannoudiji, J. Dupont-Roc, and G. Grynberg, {\it
Atom-Photon Interactions}, (Wiley, New York, 1992).

\bibitem{FordOconnell97}
G. W. Ford and R. F. O'Connell, Physica A {\bf 243}, 377 (1997).

\bibitem{LossDeVincenzo03}
D. Ahn, J. Lee, M. S. Kim, S. W. Hwang, Phys. Rev. A {\bf 66},
012302 (2002); D. Loss and D. P. DiVincenzo, ``Exact Born
Approximation for the Spin-Boson Model'', cond-mat/0304118.


\bibitem{YHCCQW02}
Y. Yu, S. Han, X. Chu, S. I. Chu, Z. Wang, Science {\bf 296}, 8898
(2002).

\bibitem{MNA02}
J. M. Martines, S. Nam, J. Aumentado, and C. Urbina, Phys. Rev.
Lett. {\bf 89}, 117901-1 (2002).

\bibitem{VACJPUED02}D. Vion, A. Aassime, A. Cottet,
P.Joyez, H. Pothier, C. Urbina, D. Esteve, M. H. Devoret, Science
{\bf 296}, 886 (2002).

\bibitem{CNHM03}
I. Chiorescu, Y. Nakamura, C. H. P. M.
Harmans, J. E. Mooij, Science {\bf 299}, 1869 (2003).

\bibitem{GargKim89} A. Garg and G. H. Kim, Phys. Rev. Lett. {\bf 63},
2512 (1989); P. M. V. B. Barone and A. O. Caldeira, Phys. Rev.
A43, 57 (1991); S. A. Egorov and B. J. Berne, J. Chem. Phys. {\bf
107}, 6050 (1997).

\bibitem {Positivity}
P. Pechukas, in {\it Large-Scale Molecular Systems}, NATO ASI
series V258, (Plenum Press, New York,1991).



\bibitem{TCL}
N. Hashitsume, F. Shibata, and M. Shingu, J. Stat. Phys. {\bf 17},
155 (1977).








\bibitem{JJ}
Y. Makhlin, G. Sch\"{o}n, A. Shnirman, Rev. Mod. Phys.
{\bf 73}, 357 (2001).

\bibitem{BXRGSJADLW03} A. J. Berkley, H. Xu, R. C. Ramos, M.
A. Gubrud, F. W. Strauch, P. R. Johnson, J. R. Anderson, A. J.
Dragt, C. J. Lobb, and F. C. Wellstood, Science {\bf 300}, 1548
(2003).

\bibitem{BKD04}
 G. Burkard, R. H. Koch, D. P. DiVincenzo,
Phys. Rev. B 69, 064503 (2004).


\bibitem {NielsenChuang}
M. A. Nielsen and I. L. Chuang, {\it Quantum Computation and
Quantum Information}, (Cambridge University Press, Cambridge,
2000).


\bibitem {RomeroPaz97}
L. D. Romero, and J. P. Paz, Phys. Rev. A {\bf 55}, 4070 (1997).


\end{thebibliography}
\end{document}